\documentclass[twocolumn,amsmath,amssymb,footinbib,superscriptaddress]{revtex4-2}

\usepackage{graphicx,color}
\usepackage{dcolumn}
\usepackage{bm}

\usepackage{epstopdf}
\usepackage{amsmath}
\usepackage[colorlinks=true,citecolor=blue,urlcolor=blue,linkcolor=blue]{hyperref}

\begin{document}

\title{Separation of interface and substrate carrier dynamics at a heterointerface\\ based on coherent optical phonons}

\author{Kunie Ishioka}
\email{ishioka.kunie@nims.go.jp}
\affiliation{National Institute for Materials Science, Tsukuba, 305-0047 Japan}

\author{Ethan Angerhoffer}
\affiliation{Department of Physics, University of Florida, Gainesville, FL 32611 USA}

\author{Christopher J. Stanton}
\affiliation{Department of Physics, University of Florida, Gainesville, FL 32611 USA}

\author{Gerson Mette}
\affiliation{Faculty of Physics and Materials Sciences Center, Philipps-Universit{\"a}t Marburg, 35032 Marburg, Germany}

\author{Kerstin Volz}
\affiliation{Faculty of Physics and Materials Sciences Center, Philipps-Universit{\"a}t Marburg, 35032 Marburg, Germany}

\author{Wolfgang Stolz}
\affiliation{Faculty of Physics and Materials Sciences Center, Philipps-Universit{\"a}t Marburg, 35032 Marburg, Germany}

\author{Ulrich H{\"o}fer}
\affiliation{Faculty of Physics and Materials Sciences Center, Philipps-Universit{\"a}t Marburg, 35032 Marburg, Germany}

\date{\today}

\begin{abstract}

Transient reflectivity spectroscopy is widely used to study ultrafast carrier- and phonon-dynamics in semiconductors. In their heterostructures, it is often not straightforward to distinguish  contributions to the signal from the various layers.  In this work,  we perform transient reflectivity measurements on lattice-matched GaP/Si(001) using a near infrared pulse, to which GaP is transparent.  The pump laser pulse can generate coherent longitudinal optical (LO) phonons both in the GaP overlayer as well as in the Si substrate which have distinct frequencies.  This enables us to track the amplitude of the respective signal contributions as a function of GaP layer thickness $d$.  The Si phonon amplitude in the signal exhibits an oscillatory behavior with increasing $d$. This can be quantitatively explained by the interference of the probe light reflected at the air/GaP/Si heterointerface.  Based on this knowledge, we can then  separate the interface- and the substrate-contributions in the carrier-induced non-oscillatory transient reflectivity signal.  The obtained interface signal provides evidence for  ultrafast carrier injection from the Si substrate into the GaP overlayer. This is also corroborated by examining the deviation of the polarization-dependence of the GaP coherent optical phonon signal  from that of the bulk semiconductor.

\end{abstract}

\maketitle

\section{INTRODUCTION}

Ultrafast carrier- and phonon-dynamics in semiconductors and their heterostructures can fundamentally influence the performance of electronic devices and have therefore been studied extensively by a variety of theoretical and experimental methods \cite{YuCardona, Sjakste2018}. 
Transient reflectivity, a pump/probe,  light-in, light-out technique based on a linear optical process, is among the most conventional and widely used experimental methods.  It is particularly powerful to detect coherent optical and acoustic phonons, which can be induced by ultrashort laser pulses and detected as periodic modulations of the transient reflectivity at THz and GHz frequencies \cite{Dekorsy, Matsuda2015}.  Because the modulation frequencies are characteristic to the materials, it is relatively straightforward to interpret phonon-induced signals obtained from heterojuncions into contributions from different semiconductor layers.   This is not the case for transient reflectivity associated with dynamics of the photoexcited carriers, which typically manifests itself  as a superposition of exponential functions.  
It requires simultaneous measurements of transient reflectivity and a non-linear spectroscopy with surface- and interface-selectivities, such as second harmonic generation (SHG), to precisely specify the carrier contributions from specific layers \cite{Glinka2008}.  

Among various semiconductor heterostructures, GaAs/AlAs quantum wells and superlattices have been studied most extensively \cite{Dekorsy1996, Yee1999, Foerst2007}.   
For other combinations of semiconductors, however, lattice mismatch often leads to a strain at the heterointerface, which can crucially affect the electronic and phononic properties \cite{Castrillo1997, Gleize1999, Jeong2016}.
Recently fabrication of abrupt GaP/Si heterointerfaces without extended defects has been made possible \cite{Thanh2012, Lin2013, Volz2011, Beyer2011, Beyer2012, Beyer2016, Belz2018, Beyer2019}. They can also serve as a model heterojunction because of the small lattice mismatch and small intermixing at the interface.  
In previous studies, systematic transient reflectivity measurements revealed the generation of coherent longitudinal optical (LO) and longitudinal acoustic (LA) phonons upon above-bandgap photoexcitation of the GaP layer and the Si substrate \cite{Ishioka2016, Ishioka2017, Ishioka2019}.
The underlying electron-phonon interaction was found to be qualitatively similar to those of the respective bulk semiconductors under the same excitation condition \cite{Hase2003, Ishioka2015}, except for the reduced LO phonon-plasmon coupling for the thinnest GaP layer examined (thickness $d=$16 nm) \cite{Ishioka2019}.
By contrast, the GaP/Si interfaces for below-bandgap excitation of GaP remains mostly unexplored, except for a time-resolved SHG study on an extremely thin ($d=$4.5 nm) GaP layer \cite{Mette2020}.   There, a fast ($<$400 fs) rise and decay in the SH signal was detected, with its intensity peaked at pump photon energy of 1.4 eV.  In addition, a delayed ($\sim$2 ps) rise was observed for pump energies above 1.4 eV.  These observations were interpreted as an electronic transition involving a short-lived electronic state at the heterointerface, whose energy lie in the bandgaps of the two semiconductors, and the subsequent transport of the photoexcited charge carriers.

In the present study we investigate the GaP/Si(001) heterointerfaces in pump-probe reflectivity scheme with near infrared optical pulses, whose photon energy exceeds the indirect bandgap of Si but is well below that of GaP.  
The reflectivity signals are periodically modulated at the well-resolved frequencies of the LO phonons of GaP and Si, which enables us to separately analyze the phonon signals from the two semiconductors.  
The amplitude of the Si coherent phonon exhibits an oscillatory dependence on the GaP overlayer thickness, which can be explained quantitatively by taking into account the interference of the probe wave reflected at the heterointerface.  This finding enables us to unambiguously decompose the carrrier-induced reflectivity signal into the the contributions from the interface and from the substrate.

\section{EXPERIMENTAL methods}

\begin{figure}
\includegraphics[width=0.475\textwidth]{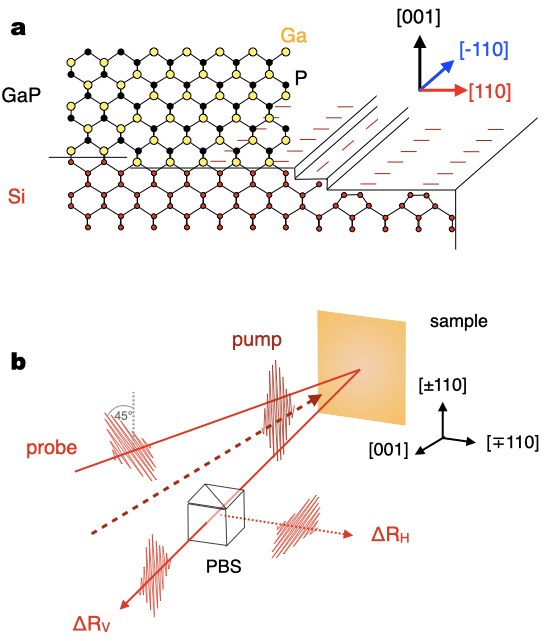}
\caption{\label{model} (a) Model of the GaP/Si(001) interface \cite{Beyer2012}.  Red, yellow and black circles represent  silicon, gallium and phosphorus atoms, respectively. (b) Schematic illustration of the configuration for the anisotropic reflectivity detection.  PBS denotes polarizing beam splitter cube.}
\end{figure}

The samples studied are nominally undoped GaP films grown by metal organic vapor phase epitaxy with thickness between $d$=8 and 48 nm on exact Si(001) substrates.  Details of the fabrication procedure are described elsewhere \cite{Volz2011, Beyer2012}.  
An 8-nm thick nucleation layer of GaP is first grown in a flow-rate modulated epitaxy at 450$^\circ$C for all the samples studied. 
The as-grown GaP nucleation layer consists of crystalline grains with lateral size of $\lesssim$30 nm \cite{Volz2011}. 
This is followed by the overgrowth of GaP in a continuous epitaxy at 675$^\circ$C for the samples $d\geq18$ nm. The GaP layers after the overgrowth have steps that directly trace those of the Si substrate \cite{Volz2011, Beyer2012}, as shown schematically in Fig.~\ref{model}a.  

\begin{figure*}
\includegraphics[width=0.9\textwidth]{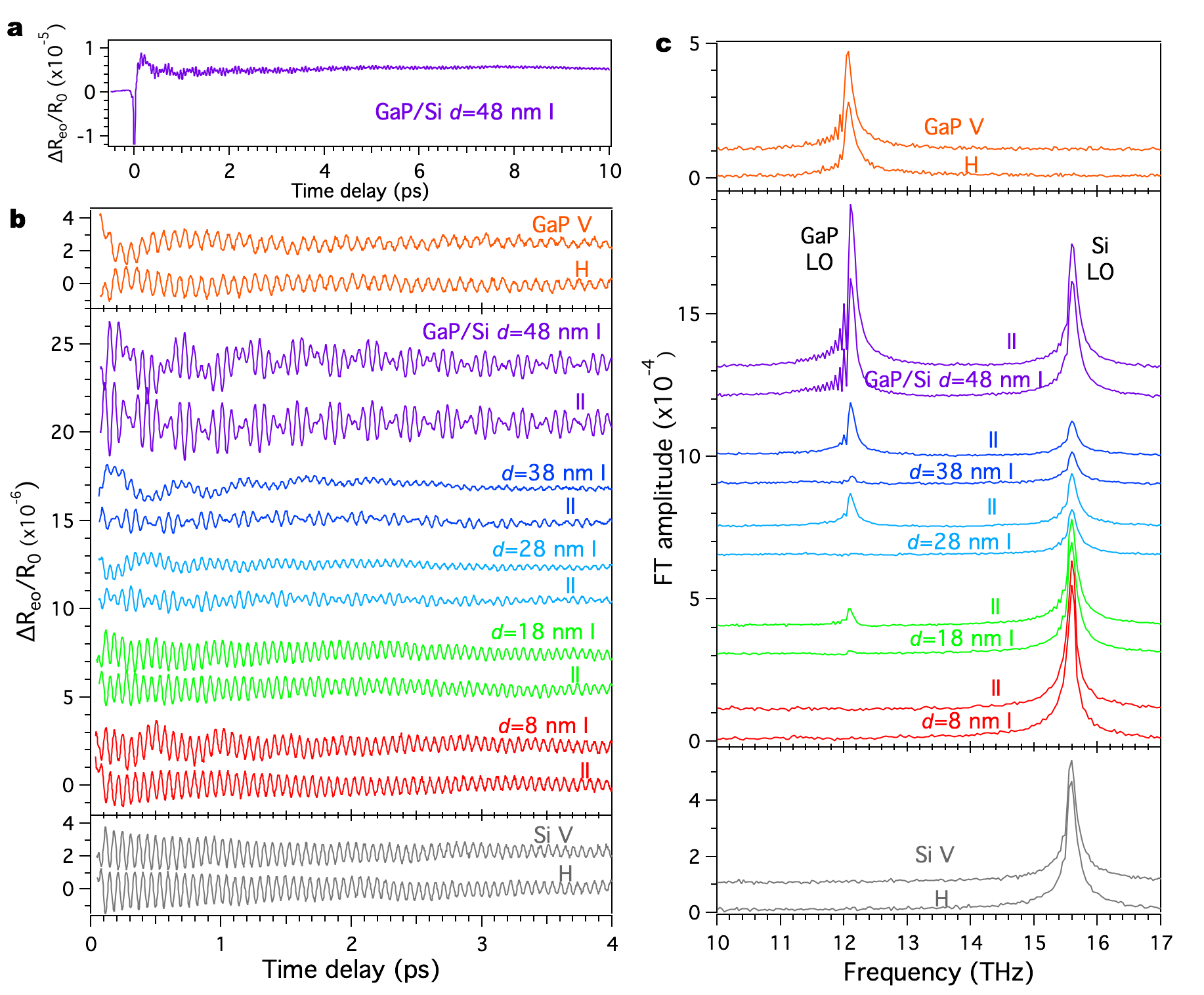}
\caption{\label{TDFTEO} (a) Anisotropic reflectivity change for $d=$48 nm. (b) Oscillatory parts of the anisotropic reflectivity changes of GaP/Si(001) samples with different $d$.  (c) Fast Fourier transform spectra of (b) in the optical phonon regime.  Pump polarization is either parallel to the [-110] or [110] directions of the Si substrate (labeled respectively with I and II).  Probe light is polarized nearly along the [100] direction for the anisotropic detection.  Reflectivity signals from (001)-oriented Si and GaP wafers with pump polarization parallel to the [-110] and [110] axes (labeled with I' and II') are also shown for comparison.  Incident pump density is 0.18 mJ/cm$^2$.  Traces are offset for clarity. 
}
\end{figure*}

Single-color pump-probe reflectivity measurements are performed in ambient conditions in a near back reflection geometry.  To investigate the phonon dynamics, an output of a Ti:sapphire oscillator with 12-fs duration, 815-nm center wavelength and 80-MHz repetition rate is used as the light source.  GaP is transparent to the 815-nm light, whereas the optical penetration depth in Si is $\sim10~\mu$m \cite{Aspnes1983}.  A spherical mirror brings the linearly polarized pump and probe beams to a $\sim$30-$\mu$m diameter spot on the sample with incident angles of $\lesssim$15$^\circ$ and $\lesssim$5$^\circ$ from the surface normal.  
Anisotropy in the pump-induced reflectivity change, $\Delta R_{eo}\equiv\Delta R_H-\Delta R_V$, is measured by detecting the horizontal ($H$) and vertical ($V$) polarization components of the probe light with a pair of matched photodiode detectors, as shown in Fig.~\ref{model}b.   This scheme is suitable to detect the LO phonons of GaP and Si, which have only off-diagonal Raman tensor components as described by eq.~(\ref{Rtensor}) in Appendix \ref{AA}, but not ideal to monitor the mostly isotropic carrier dynamics.  The signal from the detector pair is pre-amplified and is averaged in a digital oscilloscope typically over 10,000 times while the delay $t$ between the pump and probe pulses is scanned continuously with a fast scan delay.  

To examine the carrier dynamics, an output of a regenerative amplifier with 150 fs duration, 810 nm wavelength, and 100 kHz repetition rate is used as the light source.  The pump and probe spot size on the sample is $\sim100~\mu$m. Pump light is chopped at a frequency of $\sim$2 kHz for a lock-in detection.  Pump-induced change in the reflectivity $\Delta R$ is measured by detecting the probe lights before and after the reflection with a pair of matched photodiode detectors.  The signal from the detector pair is amplified with a current pre-amplifier and a lock-in amplifier.  The time delay $t$ between the pump and probe pulses is scanned with a linear motor stage (slow scan).

\section{RESULTS}


We first examine the phonon dynamics by measuring $\Delta R_{eo}/R_0$. Figure~\ref{TDFTEO}a shows a typical anisotropic reflectivity change for the $d=$48 nm GaP/Si sample.   We extract its oscillatory part by subtracting the non-oscillatory baseline that can be fitted to a multiple exponential function.  The obtained oscillations are summarized  in Fig.~\ref{TDFTEO}b for all the GaP/Si samples at two representative pump polarizations.  The oscillations consist mainly of two frequencies, 12 and 15.6 THz, which are seen as sharp peaks in the Fast Fourier-transformed (FFT) spectra in Fig.~\ref{TDFTEO}c.  These peaks arise from coherent LO phonons of GaP and Si, respectively, as is evident from the comparison with the signals of bulk GaP and Si shown in the same figure
\footnote{The oscillations shown in Fig.~\ref{TDFTEO} also include a low-frequency modulation at 2 THz, which is especially evident for the pump polarization labeled with II.  This modulation is seen only for the GaP/Si samples and absent for the bulk GaP and Si. We tentatively attribute this modulation to a phonon mode localized at the GaP/Si interface, and will report further results in a separate publication.}.

For the bulk GaP and Si crystals the [-110] and [110] crystallographic directions are equivalent.  Correspondingly, the LO phonon amplitudes are comparable between the pump polarizations along these two directions, as shown in Fig.~\ref{TDFTEO}bc, while the phases of the oscillations are opposite to each other because of the Raman generation, as explained in Appendix \ref{AA}.
For the GaP/Si samples, by contrast, the two directions can be distinguished based on the miscut of the Si substrate surface, as schematically shown in Fig.~\ref{model}a.  
The GaP peak height in Fig.~\ref{TDFTEO}c is apparently larger for pump polarization along the [110] axis  (labeled ``II") than along the [-110] axis (``I").  
For a fixed pump polarization, the GaP peak height increases monotonically with increasing $d$.  
By contrast, the Si peak height is comparable between the two polarizations for all the GaP/Si samples examined. It depends on $d$ in an apparently complicated manner, however, i.e., first decreases and then increases with increasing $d$.  

\begin{figure}
\includegraphics[width=0.475\textwidth]{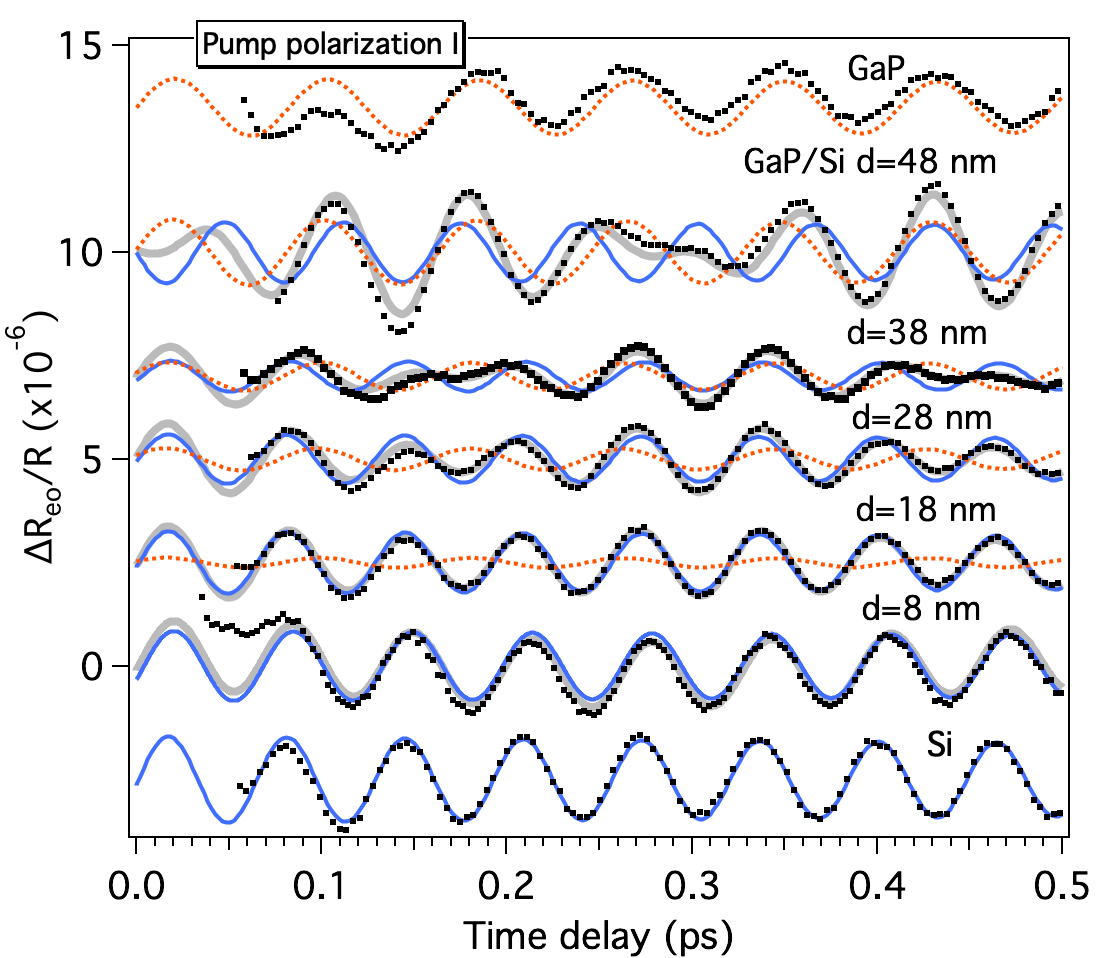}
\caption{\label{fit_components} Comparison of oscillatory parts of $\Delta R_{eo}/R_0$ obtained at pump polarization I (black dots) with fits to eq.~(\ref{dh}) (grey curves).  Blue solid and orange dotted curves represent the fit components of Si and GaP LO phonons, respectively.  Signals and fits for the bulk Si and GaP are also shown for comparison.  Traces are offset for clarity.}
\end{figure}

For quantitative analyses we fit the the oscillatory signals to a multiple damped harmonic function: 
\begin{equation}\label{dh}
f(t)=\sum_i A_i \exp(-\Gamma_i t) \sin(2\pi\nu_i t+\phi_i),
\end{equation}
with $i$ denoting different phonon modes.  Figure~\ref{fit_components} compares the experimentally obtained oscillations with the fits and their GaP and Si phonon components at a fixed pump and probe polarization combination.  Whereas the GaP oscillation component simply increases in the amplitude with increasing $d$, the Si component flips its phase between $d$=38 and 48 nm. 
We note that the frequencies $\nu_i$ and the dephasing rates $\Gamma_i$ show no systematic dependence on the GaP thickness $d$ and agree with those of the bulk GaP and Si within experimental errors. 

\begin{figure}
\includegraphics[width=0.475\textwidth]{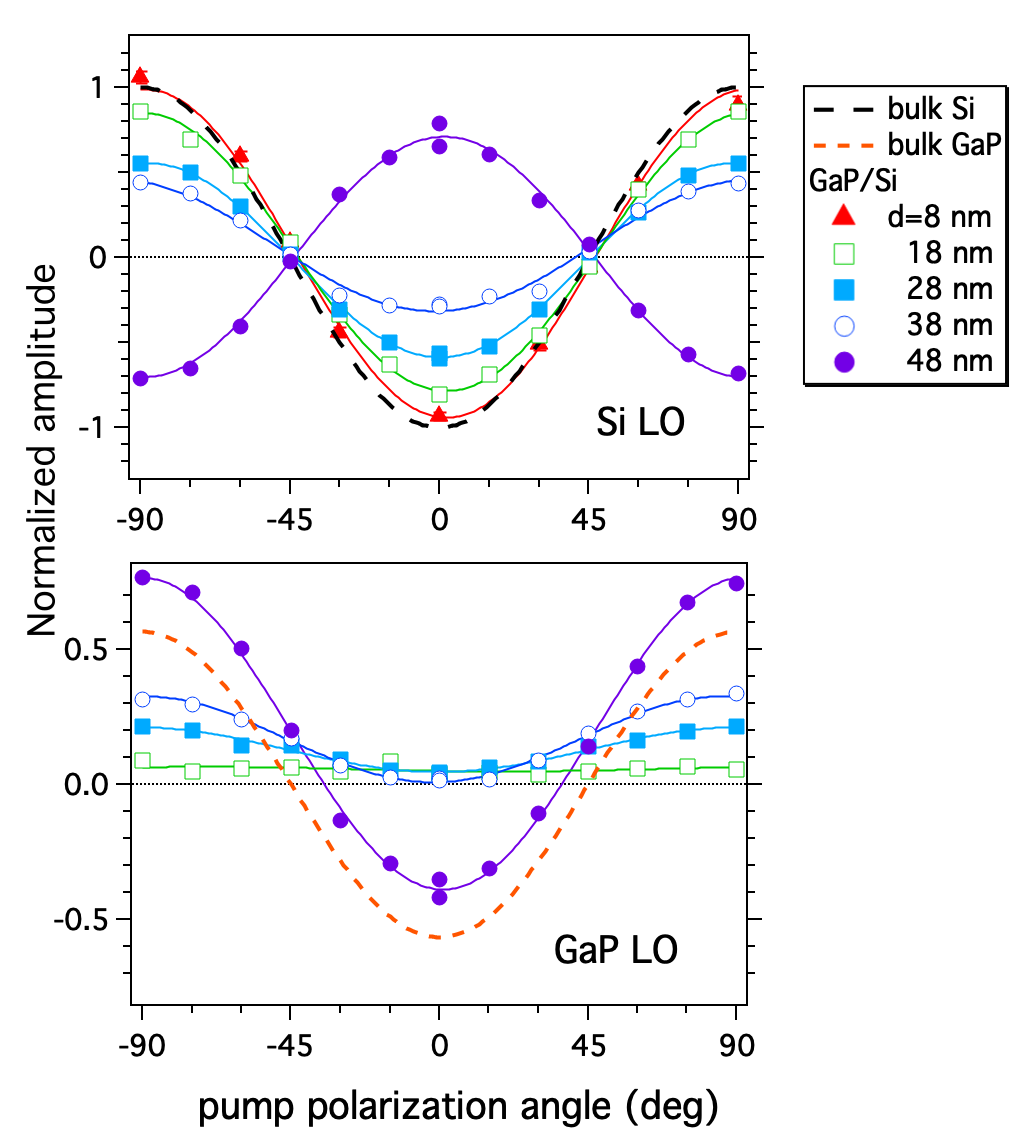}
\caption{\label{PumpPolEO} LO phonon amplitudes $A_\textrm{Si}$ (top panel) and $A_\textrm{GaP}$ (bottom panel) of the GaP/Si samples as a function of pump polarization angle $\theta$ from the [110] direction of the Si substrate.  Vertical axes are normalized by $A_\textrm{Si}$ of bulk Si at $\theta=90^\circ$.  Solid curves represent the fits to eqs.~(\ref{fit}) and (\ref{fit2}).  Amplitudes for the bulk Si and GaP are also shown with broken curves for comparison.}
\end{figure}

Figure~\ref{PumpPolEO} plots the GaP and Si phonon amplitudes, $A_\text{GaP}$ and $A_\text{Si}$, as a function of pump polarization angle $\theta$  from the [110] axis of the Si substrate.
Here we restrict the initial phase $\phi_i$ around zero and allow $A_i$ to take a positive or negative value to represent the phase flip.    
The Si phonon amplitude (top panel of Fig.~\ref{PumpPolEO})  always follows a cosine function of $\theta$:
\begin{equation}\label{fit}
A_\textrm{Si}(d, \theta)= -B_\textrm{Si}(d)\cos2\theta,
\end{equation}  
for all the GaP/Si samples as well as for the bulk Si.  This is the manifest of the generation of the coherent LO phonons via impulsive stimulated Raman scattering (ISRS), as described by eq.~(\ref{force2}) in Appendix \ref{AA}.
The GaP phonon amplitude for the \emph{bulk} GaP, plotted with an orange broken curve in the bottom panel of Fig.~\ref{PumpPolEO}, shows a similar $\theta$-dependence, indicating an ISRS-generation in the present below-bandgap excitation condition \cite{Ishioka2015}.   

The GaP phonon amplitude for the GaP/Si samples, by contrast, is described more appropriately with an additional $\theta$-independent term by: 
\begin{equation}\label{fit2}
A_\textrm{GaP}(d,\theta)=C_\textrm{GaP}(d)- B_\textrm{GaP}(d)\cos2\theta.
\end{equation}
A similar $\theta$-independent contribution was also observed for \emph{bulk} GaP upon above-bandgap photoexcitation, and was attributed to the ultrafast screening of the dc field in the surface depletion region by photoexcited carriers  \cite{Ishioka2015}.  In a cubic crystal such as GaP, the driving force for this transient depletion field screening (TDFS) mechanism is independent of the pump polarization, as described by eq.~(\ref{TDFSforce}) in Appendix \ref{AA}.  
Accordingly, we attribute the $\theta$-independent term $C_\text{GaP}$ of the GaP/Si samples to the TDFS-driven LO phonons.  Because the phonon amplitude $A_\text{GaP}$ increases linearly with increasing pump fluence for all the GaP/Si samples (not shown in Figure), it is unlikely that the photocarriers are created via a two-photon transition within the GaP layer.  The more likely source of the charge carrier is via the injection at the heterointerface, either from the Si substrate or from the interface electronic states reported in the previous SHG study \cite{Mette2020}.  

\begin{figure}
\includegraphics[width=0.475\textwidth]{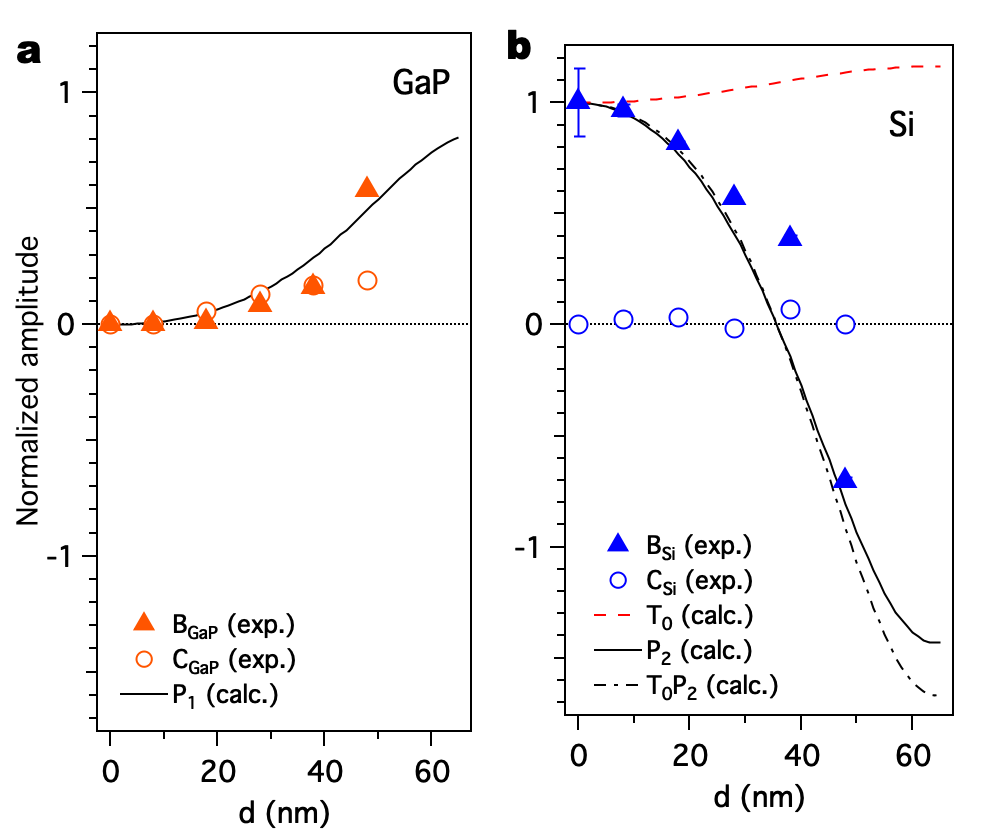}
\caption{\label{A_vs_d}  $\theta$-dependent ($B$, filled triangles) and $\theta$-independent ($C$, empty circles) amplitude components of GaP (a) and Si (b) LO phonons, as a function of GaP layer thickness $d$.  The components are normalized by $B_\textrm{Si}$ of bulk Si ($d=0$ nm).  
Solid curves in (a) and (b) represent $P_1$ and $P_2$ calculated with eqs.~(\ref{omega1}) and (\ref{omega2}).  Red broken and black chain curves in (b) represent $T_0$ given by eq.~(\ref{D2}) and $T_0P_2$, respectively.   $P_1$ and $P_2$ are normalized by $P_2(d=0)$, whereas $T_0$ is normalized by $T_0(d=0)$.
}
\end{figure}

Figure~\ref{A_vs_d} summarizes the $d$-dependence of the amplitude components $B_i$ and $C_i$ obtained from fitting $A_i(\theta)$  to eq.~(\ref{fit2}).  The components for the GaP LO mode, $B_\textrm{GaP}$ and $C_\textrm{GaP}$, increase in parallel with increasing $d$ up to 38 nm.  At $d$=48 nm, $B_\textrm{GaP}$ becomes several times larger than $C_\textrm{GaP}$, indicating that the bulk-like ISRS overwhelms the interface-specific TDFS for the thickest GaP layer.  
By contrast, the component for the Si LO mode, $B_\text{Si}$, decreases almost monotonically with increasing $d$ up to 38 nm.  At $d$=48 nm it takes a negative value, denoting a phase flip of the coherent oscillation. 
We will discuss the origin of this peculiar behavior in Sect.~\ref{IV}.


\begin{figure}
\includegraphics[width=0.475\textwidth]{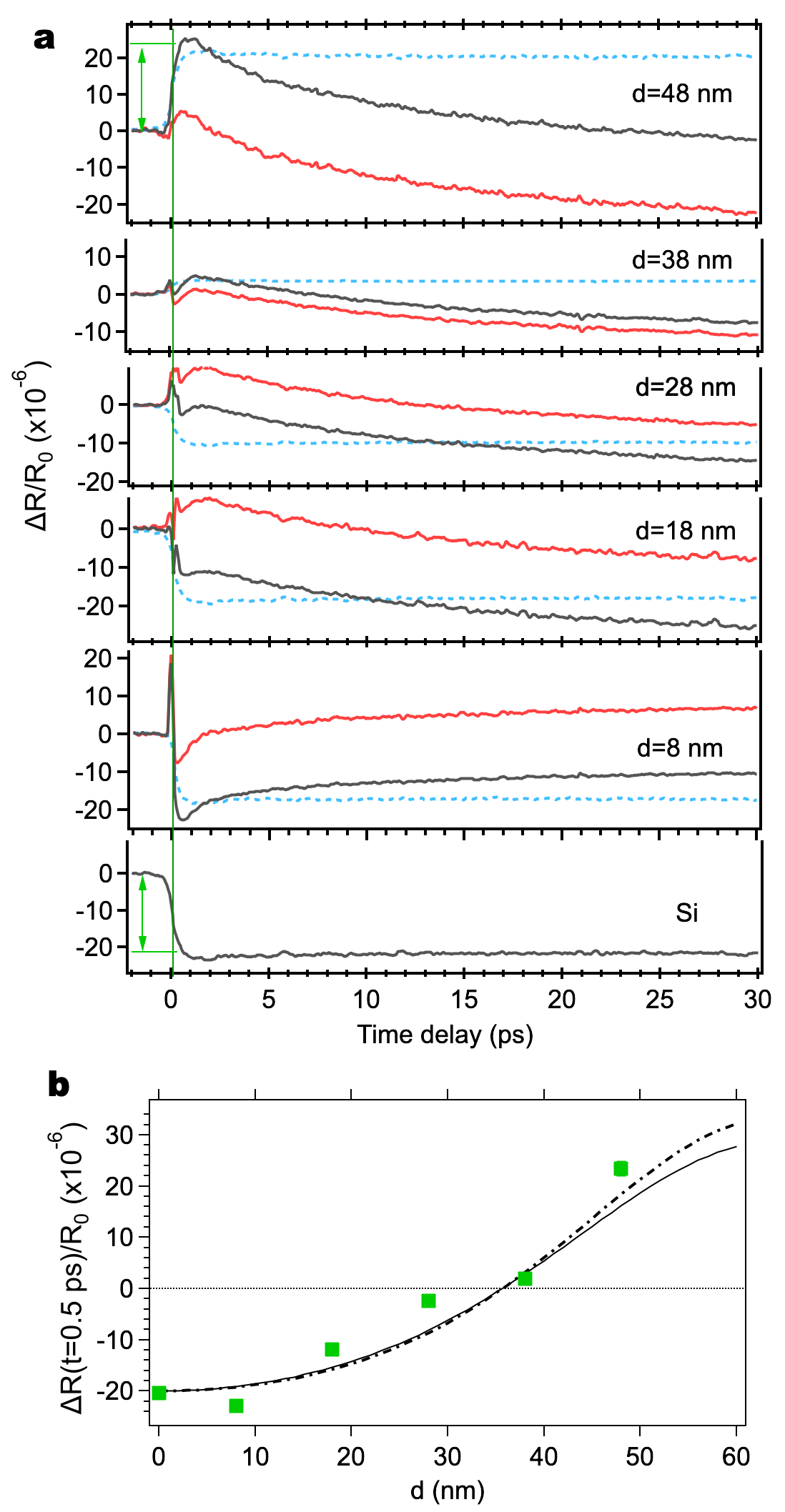}
\caption{\label{Slow} (a) As-measured transient reflectivity signals $\Delta R/R_0$ (grey curves) of GaP/Si(001) and bulk Si.  Pump and probe lights are polarized along the [110] axes of the Si substrate.  Incident pump density is 0.25 mJ/cm$^2$.  Vertical line indicates $t=0.5$ ps, and arrows indicate the initial step height $\Delta R(t=0.5 \text{ps})/R_0$ for selected traces.  Blue broken and red solid curves indicate the contributions from the substrate and the interface given by eqs.~(\ref{Si_sub}) and (\ref{Si_sub2}).  (b) Initial step height $\Delta R(t=0.5 \text{ps})/R_0$ as a function of GaP layer thickness $d$.  
Solid and chain curves represent the $d$-dependence of $P_2$ and $T_0P_2$.   Curves are scaled to $\Delta R(t=0.5 \text{ps})/R_0$ of bulk Si.
}
\end{figure}

In order to examine the carrier dynamics that can possibly contribute to the TDFS, we also measure the transient reflectivity $\Delta R/R_0$. 
Black curves in  Fig.~\ref{Slow}a compare the as-measured reflectivity changes of the GaP/Si samples  with that of bulk Si.  The signal from the bulk Si shows a step-function-like drop at $t=$0, followed by very little recovery in the present time window.  The response is in good agreement with previous reports \cite{Sabbah2000, Sabbah2002} and can be attributed to the free carrier excitation across the indirect bandgap, followed by their diffusion and recombination at the surface.
Bulk GaP, by contrast, exhibits no detectable change under the present photoexcitation condition and its signal is therefore not shown.

Transient reflectivity traces of the GaP/Si samples appear in qualitative contrast to those of the bulk Si and GaP, indicating the carrier dynamics that is characteristic to the GaP/Si heterointerface.  For $d$=8 nm, the reflectivity signal shows an abrupt drop at $t\simeq0$ that is somewhat similar to the bulk Si, though the subsequent recovery is more distinct. 
The signals from the thicker ($d\geq$18 nm) GaP/Si, by contrast, show an initial abrupt drop or rise depending on $d$, followed by a sub-picosecond increase and then a slower decrease.  
The height of the initial abrupt drop or rise, which we represent with the transient reflectivity at $t=0.5$ ps, exhibits a peculiar $d$-dependence, as shown in Fig.~\ref{Slow}b.  It starts from a negative value and increases monotonically with increasing $d$ until it reaches a positive value.  This trend is very similar to that of the Si LO phonon, $B_\text{Si}$, if we normalize the initial step height by that of the bulk Si.  

\section{discussion}\label{IV}

\begin{figure}
\includegraphics[width=0.425\textwidth]{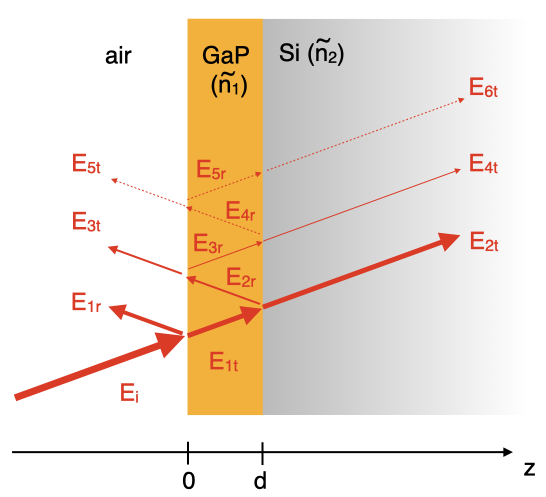}
\caption{\label{Probe}  Schematic illustration of the transmission and reflection of probe electric field that is incident on a GaP/Si interface.  $E_i$: incident wave, $E_{jr}$ and $E_{jt}$: reflected and transmitted wave at $j$-th interface.  Incidence angle is exaggerated for clarity.}
\end{figure}

In the previous Section we have seen that the phonon- and the carrier-induced signals from the Si substrate exhibit similar oscillatory behaviors as a function of overlayer thickness.  The observations hint at the involvement of the oscillating electric field of the laser pulse, which has a wavelength of $\sim$200 nm in the semiconductors.  In this Section we theoretically model the effect of the GaP overlayer on the pump-induced reflectivity signal from the buried Si substrate by explicitly taking into account the light carrier wave.

We first examine the effect of the GaP overlayer on the pump light incoming to the Si substrate, whose details are described in Appendix~\ref{AB}.  We assume a light pulse, whose electric field in air is described by:
\begin{eqnarray}\label{D1}
E_{i}(z,t)&=&\tilde{E}_i\left(z,t\right)e^{i(k_0 z-\omega_0 t)},
\end{eqnarray}
being incident on a GaP/Si heterointerface from the normal direction.
At the air/GaP ($z=0$) and GaP/Si ($z=d$) interfaces, the light pulse is partially reflected and partially transmitted, as schematically shown in Fig.~\ref{Probe}.  
Taking into the multiple reflections, the transmittance of the heterointerface, or the intensity ratio of the light penetrating into Si to the incident light, can be expressed by:
\begin{eqnarray}\label{D2}
T_0(d)&=&\dfrac{(1-r_{01}^2)(1-r_{12}^2)}{1+r_{01}^2r_{12}^2+2r_{01}r_{12}\cos{2n_1k_0d}},
\end{eqnarray}
with 
\begin{eqnarray}\label{D3}
r_{01}&\equiv&\dfrac{\tilde{E}_{1r}}{\tilde{E}_i}=\dfrac{1-{n}_1}{1+{n}_1},\nonumber\\
r_{12}&=&\dfrac{\tilde{E}_{2r}}{\tilde{E}_{1t}}=\dfrac{{n}_1-{n}_2}{{n}_1+{n}_2}
\end{eqnarray}
representing the reflection coefficients for the light wave coming from air into GaP and from GaP into Si, respectively.  ${n}_1$ and ${n}_2$ are the refractive indices of GaP and Si. 

Equation~(\ref{D2}) implies that the pump \emph{intensity} in the Si substrate can depend on the overlayer thickness $d$ due to the interference among different optical paths ($E_{2t}, E_{4t}, E_{6t}, \cdots$ in Fig.~\ref{Probe}).  This is the same principle as used for anti-reflection coatings on optics surfaces.  In the present case of GaP/Si, however, the contribution from the $d$-dependent term is insignificant, because $r_{12}$ is small as a consequence of comparable refractive indices of GaP ($n_1$=3.18) and Si ($n_2$=3.68).  The pump intensity in Si is modified no more than by $15\%$ by varying $d$, as shown with a red broken curve in Fig.~\ref{A_vs_d}b, which by itself cannot explain the drastic thickness-dependence of the Si phonon and carrier signals observed in our experiments.

Next we examine the effect of the GaP overlayer on the probe light, whose detail is described in Appendix \ref{AC}. The overlayer similarly modifies the reflectance of the interface: 
\begin{eqnarray}\label{D0}
R_0(d)&=&\dfrac{r_{01}^2+r_{12}^2+2r_{01}r_{12}\cos2n_1k_0d}{1+r_{01}^2r_{12}^2+2r_{01}r_{12}\cos{2n_1k_0d}}.
\end{eqnarray}
We assume a separate pump pulse induces small disturbances $\delta{n}_1(z,t)$ and $\delta{n}_2(z,t)$ in the refractive indices of the GaP layer and the Si substrate. 
Coherent phonons in the respective semiconductors modulate the refractive indices periodically as a function of $t$, whereas photoexcited carriers in Si induces a step-function-like change in ${n}_2$.
In either case, the transient reflectivity can be expressed by taking its derivatives with respect to the refractive indices:
\begin{equation}\label{D4}
\dfrac{\Delta R}{R_0}=\dfrac{1}{R_0}\left(\dfrac{\partial R_0}{\partial {n}_1}\delta {n}_1+\dfrac{\partial R_0}{\partial {n}_2}\delta {n}_2\right)\equiv P_1\delta n_1+P_2 \delta n_2.
\end{equation}
The first and second terms of eq.~(\ref{D4}) represent the pump-induced reflectivity changes contributed by the GaP overlayer and by the Si substrate.  They can be given respectively by:
\begin{widetext}
\begin{eqnarray}\label{omega1}
P_1(d)\delta n_1&=&\dfrac{1}{R_0}\left(\dfrac{\partial R_0}{\partial r_{01}}\dfrac{\partial r_{01}}{\partial n_1}+\dfrac{\partial R_0}{\partial r_{12}}\dfrac{\partial r_{12}}{\partial n_1}+\dfrac{\partial R_0}{\partial (n_1k_0d)}\dfrac{\partial (n_1k_0d)}{\partial n_1}\right)\delta n_1\nonumber\\
&=&\dfrac{2(r_{01}^2-1)(1-r_{12}^2)}{(r_{01}^2+r_{12}^2+2r_{01}r_{12}\cos2n_1k_0d)(1+r_{01}^2r_{12}^2+2r_{01}r_{12}\cos{2n_1k_0d})}\nonumber\\
&\times&\left[\dfrac{r_{01}(1+r_{12}^2)-r_{12}(1+r_{01}^2)}{2n_1}(1-\cos{2n_1k_0d})+2k_0dr_{01}r_{12}\sin2n_1k_0d\right]\delta n_1,
\end{eqnarray}
\begin{eqnarray}\label{omega2}
P_2(d)\delta n_2&=&\dfrac{1}{R_0}\dfrac{\partial R_0}{\partial r_{12}}\dfrac{\partial r_{12}}{\partial n_2}\delta n_2\nonumber\\
&=&\dfrac{2r_{12}(1-r_{01}^4)+2r_{01}(1+r_{12}^2)(1-r_{01}^2)\cos2(n_1k_0d)}{(r_{01}^2+r_{12}^2+2r_{01}r_{12}\cos2n_1k_0d)(1+r_{01}^2r_{12}^2+2r_{01}r_{12}\cos2n_1k_0d)}\times\dfrac{-2n_1}{(n_1+n_2)^2}\delta n_2.
\end{eqnarray}
\end{widetext}

Solid curves in Fig.~\ref{A_vs_d} show the $d$-dependences of the pump-induced reflectivity changes given by eqs.~(\ref{omega1}) and (\ref{omega2}). The calculations reasonably reproduce the experimentally obtained Si and GaP phonon signals, $B_\text{Si}$ and $B_\text{GaP}$, of the GaP/Si samples.
Equation~(\ref{omega2}) also reproduces the $d$-dependence of the initial step height, $\Delta R(t=0.5 \text{ps})/R_0$, as shown in Fig.~\ref{Slow}b, indicating that the initial drop/rise arise mostly (but not entirely) from the photoexcitation in the Si substrate.  Taking into account the $d$-dependence of the pump intensity (eq.~(\ref{D2})) introduces only a minor correction, as shown with chain curves in Figs.~\ref{A_vs_d}b and \ref{Slow}b.  The agreements confirm that the interference effect on the reflected \emph{probe} light is the origin of the oscillatory behavior of the transient reflectivity signal from the Si substrate, whether it is associated with phonons or carriers.  

We can now use eq.~(\ref{omega2}) to decompose the carrier-induced signal into the substrate- and interface-contributions.  We can express the Si substrate contribution with:   
\begin{eqnarray}\label{Si_sub}
\dfrac{\Delta R_\text{sub}(t,d)}{R_0}&=&\dfrac{T_0(d)P_2(d)}{T_0(0)P_2(0)}\dfrac{\Delta R_\text{Si}(t)}{R_0},
\end{eqnarray}
where $\Delta R_\text{Si}(t)/R_0$ represents the transient reflectivity signal of the bulk Si obtained under the same condition. The contribution from the interface is then given by:
\begin{eqnarray}\label{Si_sub2}
\dfrac{\Delta R_\text{int}(t,d)}{R_0}&=&\dfrac{\Delta R(t,d)}{R_0}-\dfrac{\Delta R_\text{sub}(t,d)}{R_0}.
\end{eqnarray}
Blue broken and red solid curves in Fig.~\ref{Slow}a show the substrate- and interface-contributions given by eqs.~(\ref{Si_sub}) and (\ref{Si_sub2}), respectively.  
We see that $\Delta R_\text{int}(t,d)/R_0$ for $d=8$ nm exhibits an abrupt drop at $t\sim0$, followed by a bi-exponential increase with time constants $\tau_\text{fast}$=1.6 ps and $\tau_\text{slow}$=43 ps toward a positive value.  The slow time constant is comparable to that of the bulk Si, 38 ps, and can be interpreted in the similar relaxation processes involving the carrier diffusion within the Si substrate and their recombination at the GaP/Si interface. The larger amplitude of this slow relaxation indicates higher density of the recombination centers at the GaP/Si heterointerface than at the naturally oxidized surface of the bulk Si.  
The fast time constant is close to the delayed rise time observed in the previous SHG study \cite{Mette2020} and may be associated with the carrier injection from the occupied interface electronic state. 

For the overgrown GaP layers ($d\geq18$ nm), $\Delta R_\text{int}(t,d)/R_0$ consists of a sub-picosecond rise followed by a monotonic decrease on $\sim$30 ps time scale.  
The initial rise can be attributed to the carrier injection into the GaP overlayer, which would lead to the sudden screening of the dc electric field and thereby TDFS generation of the coherent phonons.  The injection should occur sufficiently fast  ($<50$ fs) in order to give a driving force for the coherent phonons.  The time scale of the following slow decrease is independent of $d$ and comparable to those of $d=8$ nm GaP/Si and bulk the Si, suggesting a similar recombination process at the heterointerface.   
 
The present study alone cannot give a conclusive evidence for the dominant source of the carriers injected into the GaP layer.  On one hand, an early theoretical study on (001)-oriented Si/GaP interface predicted a type-I band alignment with the valence band offset $\Delta E_v$=0.8 eV \cite{Dandrea1990}.  Photoemission experiments \cite{Perfetti1984, Sakata2008} found comparable or larger $\Delta E_v$.  This would give sufficient excess energy for the carriers photoexcited in Si with a 1.5-eV light to overcome the conduction band offset and to be injected into GaP.  On the other, the previous SHG study \cite{Mette2020} indicated the occupied interface electronic state whose energy is close to the valence band maximum of Si.  The transition from the interface state exhibited a resonance peak at 1.4 eV, which is not far from the photon energy employed in the present study.  To unambiguously determine the injection pathways we are performing further transient reflectivity experiments on the GaP/Si heterointerfaces with a tunable pump light source, which will be reported in a separate publication.  

\section{conclusion}

We investigated the carrier- and phonon-dynamics of the lattice-matched GaP/Si(001) interfaces upon below-bandgap photoexcitation of GaP by means of pump-probe reflectivity measurements.   We demonstrated the contribution of a buried Si substrate to the transient reflectivity signal, whether it is of electronic or phononic origin, to exhibit an oscillatory dependence on the thickness of an optically transparent GaP overlayer.  The oscillatory behavior was quantitatively explained in terms of the interference of the probe light reflected at the heterointerface, whereas the interference effect on the pump intensity penetrating into the Si substrate was found to be minor in the present case.  Based on this finding we extracted the interface contribution in the {\it carrier-induced}  transient reflectivity traces.  The obtained signals clearly indicated ultrafast injection of charge carriers into the overlayer, which was in agreement with the polarization-dependence of the coherent LO phonon of GaP.  The knowledge obtained in the present study can be also applied to the quantitative analyses of transient reflectivity signals for wide varieties of buried semiconductor heterointerfaces. 
	
\begin{acknowledgments}
We gratefully acknowledge funding by the Deutsche Forschungsgemeinschaft (DFG, German Research
Foundation), Project-ID 223848855-SFB 1083. CJS was partially supported by the Air Force Office of Scientific Research under Award No. FA9550-17-1-0341.
\end{acknowledgments}

\bibliographystyle{apsrev4-2}
\bibliography{GaPSi800nmLO}

\appendix

\section{Coherent phonon generation and detection in bulk Si and GaP}\label{AA}

Illumination with a femtosecond laser pulse can induce coherent optical phonons in semiconductors.  The generation mechanism can depend on the semiconductor and the excitation photon energy.  
When the photon energy is below the fundamental bandgap, or when the semiconductor is transparent to the excitation light,  impulsive stimulated Raman scattering (ISRS) \cite{Dhar1994} is the only known generation mechanism.  
In ISRS, a broadband femtosecond optical pulse offers multiple pairs of photons required for the stimulated process.  The driving force $F$ depends on the polarization of the optical electric field $E$ through a third-rank Raman tensor $\mathfrak{R}_{jkl}\equiv(\partial\chi/\partial Q)_{jkl}$ \cite{Dekorsy}:
\begin{equation}\label{force}
F^\textrm{ISRS}_j(t)=\mathfrak{R}_{jkl}E_k(t)E_l(t),
\end{equation}
where $j, k, l$ denote the Cartesian coordinates. 
The Raman tensor of diamond- and zinc blende-structured crystals is given in the form of  \cite{YuCardona}:
\begin{equation}\label{Rtensor}
\mathfrak{R}_{xkl}= \begin{pmatrix}
0 & 0 & 0\\
0 & 0 & a \\
0 & a & 0
\end{pmatrix},
\mathfrak{R}_{ykl}= \begin{pmatrix}
0 & 0 & a\\
0 & 0 & 0 \\
a & 0 & 0
 \end{pmatrix},
\mathfrak{R}_{zkl}= \begin{pmatrix}
0 & a & 0\\
a & 0 & 0 \\
0 & 0 & 0
 \end{pmatrix}
\end{equation}
 In the back-reflection from the (001)-oriented surface, in which the pump light polarization has no $z$ component, the driving force can be reduced to:
\begin{eqnarray}\label{force2}
F^\textrm{ISRS}_z(t)&=&\mathfrak{R}_{zxy}E_x(t)E_y(t)+\mathfrak{R}_{zyx}E_y(t)E_x(t)\nonumber\\
&=&a|E(t)|^2\sin 2\theta'={a}|E(t)|^2\cos 2\theta,
\end{eqnarray}
with $E_x\equiv|E|\cos\theta', E_y\equiv|E|\sin\theta'$, and $\theta'\equiv\theta-\pi/4$ being the polarization angle from the [100] axis.  The driving force becomes maximum at $\theta=0$ or $\pi$, i.e., when the pump light is polarized along the [110] or [-110]  axis of the crystal.  The direction of the driving force reverses between these two polarizations, which explains the phase flip of the coherent LO phonons of bulk GaP and Si shown in Fig.~\ref{TDFTEO}b.

When the photon energy exceeds the bandgap, or when the semiconductor is opaque, 
 the ISRS generation of coherent phonons can be resonantly enhanced in the similar manner as in spontaneous Raman scattering  \cite{Stevens2002, Cardona1982}.  
 In the case of a polar semiconductor such as GaP, transient depletion field screening (TDFS) can contribute in addition \cite{Dekorsy, Ishioka2015}.  In the TDFS mechanism, separation of photoexcited electrons and holes in the surface depletion layer induces ultrafast drift-diffusion current $J_z$ in the surface normal direction and thereby offers a driving force for the coherent polar phonons \cite{Dekorsy}:
\begin{equation}\label{TDFSforce}
F^\textrm{TDFS}_z(t)=-\frac{e^*}{\varepsilon_\infty\varepsilon_0}\int_{-\infty}^t dt'J_z(t').
\end{equation}
In a cubic crystal whose optical absorption is isotropic within the \{001\} plane, the driving force is independent of the pump polarization.

The coherent phonons can be observed as a periodic modulation of reflectivity at the zone-center optical phonon frequency.  
A nuclear displacement $Q$ associated with the LO phonon oscillation induces a change in reflectivity $R$ through the refractive index $n$ and the susceptibility $\chi$.  In a first-order approximation the change $\Delta R$ is given by \cite{Dekorsy}:
\begin{equation}\label{reflectivity}
\Delta R=\frac{\partial R}{\partial n}\Delta n \simeq \frac{\partial R}{\partial \chi}\frac{\partial \chi}{\partial Q} \Delta Q.
\end{equation}
Here $\partial \chi/\partial Q=\mathfrak{R}$ is the first-order Raman tensor given in eq.~(\ref{Rtensor}). Equation (\ref{reflectivity}) implies that only Raman active phonons can be detected in transient reflectivity, and that the phonon signal depends on the probe light polarization angle $\theta$ in the same manner as the pump light described in eq.~(\ref{force2}).

\section{Reflection and transmission of light at heterointerface}\label{AB}

In this Section we consider the transmission and reflection of light at a GaP/Si heterointerface, which consists of a GaP layer of thickness $d$ on top of a semi-infinitely thick Si substrate, as schematically illustrated in Fig.~\ref{Probe}. We assume a light pulse, whose electric field in air is expressed by:
\begin{eqnarray}\label{th1}
E_{i}(z,t)&=&\tilde{E}_{i}\left(z,t\right)e^{i(k_0 z-\omega_0 t)},
\end{eqnarray}
is incident from the normal direction.
Here $\tilde{E}_i(z,t)$ and $\exp[i(k_0z-\omega_0 t)]$ represent the slowly varying envelope function and the fast varying carrier wave.  $z$ and $t$ represent the distance from the air/GaP interface and time.  $\omega_0$ and $k_0=\omega_0/c$ denote the light wave frequency and the wavevector.  Hereafter we approximate the slowly varying envelope function in eq.~(\ref{th1}) as a time-independent constant.  

At the air/GaP interface at $z=0$, the light pulse is partially reflected and partially transmitted into the GaP layer.  
The reflected and transmitted light waves can be expressed by:
\begin{eqnarray}\label{th2}
E_{1r}(z,t)&=&\tilde{E}_{1r}\left(z,t\right)e^{-i(k_0 z+\omega_0 t)},\nonumber\\
E_{1t}(z,t)&=&\tilde{E}_{1t}\left(z,t\right)e^{i(\tilde{n}_1k_0 z-\omega_0 t)},
\end{eqnarray}
where $\tilde{n}_1=n_1+i\kappa_1$ is the refractive index of GaP.  We use $n_1$=3.18 and $\kappa_1$=0 at wavelength $\lambda_0=2\pi/k_0$=815 nm \cite{Aspnes1983} in the following calculations.
We apply the boundary condition that the in-plane components of the electric and magnetic fields are continuous, and obtain the reflection and transmission coefficients for the light wave incoming from air into GaP:
\begin{eqnarray}\label{th3}
r_{01}&\equiv&\dfrac{\tilde{E}_{1r}}{\tilde{E}_i}=\dfrac{1-{n}_1}{1+{n}_1}=-0.52,\nonumber\\
t_{01}&\equiv&\dfrac{\tilde{E}_{1t}}{\tilde{E}_i}=\dfrac{2}{1+{n}_1}=0.48.
\end{eqnarray}

The light wave transmitted into the GaP layer is again partially reflected at the GaP/Si interface and partially transmitted into the Si substrate.  We describe the reflected and transmitted waves by:
\begin{eqnarray}\label{th4}
E_{2r}(z,t)&=&\tilde{E}_{2r}\left(z,t\right)e^{-i({n}_1k_0 (z{-2}d)+\omega_0 t)},\nonumber\\
E_{2t}(z,t)&=&\tilde{E}_{2t}\left(z,t\right)e^{i({n}_1k_0 d+\tilde{n}_2k_0 (z-d)-\omega_0 t)},
\end{eqnarray}
where $\tilde{n}_2=n_2+i\kappa_2$ is the refractive index of Si, with $n_2$=3.68 and $\kappa_2$=0.006 at $\lambda_0$=815 nm \cite{Aspnes1983}.  Equation~(\ref{th4}) implies that the light wave gains a thickness-dependent phase shift, {$\phi\equiv{n}_1k_0d$}, while it crosses the GaP layer once. 
We apply the boundary condition at the GaP/Si interface and obtain the reflection and transmission coefficients for the light wave incoming from GaP into Si:
\begin{eqnarray}\label{th5}
r_{12}&=&\dfrac{\tilde{E}_{2r}}{\tilde{E}_{1t}}=\dfrac{{n}_1-{n}_2}{{n}_1+{n}_2}=-0.073,\nonumber\\
t_{12}&=&\dfrac{\tilde{E}_{2t}}{\tilde{E}_{1t}}=\dfrac{2{n}_1}{{n}_1+{n}_2}=0.93.
\end{eqnarray}
Here we neglect the small optical absorption in Si and approximate $\tilde{n}_2\simeq n_2$. 
After the reflection at the GaP/Si interface, the light wave is again partially reflected at the GaP/air interface and partially transmitted into air.  We describe the reflected and transmitted waves by:
\begin{eqnarray}\label{th6}
E_{3r}(z,t)&=&\tilde{E}_{3r}\left(z,t\right)e^{i(\tilde{n}_1k_0 z-\omega_0 t{+2\phi})},\nonumber\\
E_{3t}(z,t)&=&\tilde{E}_{3t}\left(z,t\right)e^{-i(k_0 z+\omega_0 t+{2\phi})}.
\end{eqnarray}
We apply the similar boundary condition and obtain the reflection and transmission coefficients for outgoing wave from GaP into air:
\begin{eqnarray}\label{th7}
r_{10}&=&\dfrac{\tilde{E}_{3r}}{\tilde{E}_{2r}}=\dfrac{{n}_1-1}{{n}_1+1}=0.52,\nonumber\\
t_{10}&=&\dfrac{\tilde{E}_{3t}}{\tilde{E}_{2r}}=\dfrac{2{n}_1}{{n}_1+1}=1.52.
\end{eqnarray}

The amplitude ratio $r_0$ of the outgoing wave into air to the incident wave can be given by the sum of multiple reflection pathways: 
\begin{eqnarray}\label{th8}
r_0&\equiv&\dfrac{\tilde{E}_{1r}+\tilde{E}_{3t}{e^{2i\phi}+\tilde{E}_{5t}e^{4i\phi}+\tilde{E}_{7t}e^{6i\phi}+\cdots}}{\tilde{E}_{i}}\nonumber\\
&=&r_{01}+t_{01}r_{12}t_{10}{e^{2i\phi}\left(1+r_{10}r_{12}e^{2i\phi}+r_{10}^2r_{12}^2e^{4i\phi}+\cdots\right)}\nonumber\\
&=&r_{01}+\dfrac{t_{01}r_{12}t_{10}e^{2i\phi}}{1-r_{10}r_{12}e^{2i\phi}}=\dfrac{r_{01}+r_{12}e^{2i\phi}}{1+r_{01}r_{12}e^{2i\phi}}
\end{eqnarray}
Here we use the relations $t_{01}t_{10}=1+r_{10}r_{01}$ and $r_{10}=-r_{01}$ derived from eqs.~(\ref{th3}) and (\ref{th7}).
Likewise, the amplitude ratio $t_0$ of the incoming wave into Si to the incident wave can be given by:
\begin{eqnarray}\label{th9}
t_0&\equiv&\dfrac{\tilde{E}_{2t}{e^{i\phi}+\tilde{E}_{4t}{e^{3i\phi}+\tilde{E}_{6t}e^{5i\phi}+\cdots}}}{\tilde{E}_{i}}\nonumber\\
&=&t_{01}t_{12}{e^{i\phi}\left(1+r_{10}r_{12}e^{2i\phi}+r_{10}^2r_{12}^2e^{4i\phi}+\cdots\right)}\nonumber\\
&=&\dfrac{t_{01}t_{12}e^{i\phi}}{1-r_{10}r_{12}e^{2i\phi}}
\end{eqnarray}
The reflectance, or the light \emph{intensity} reflected into air, can then be given by:  
\begin{eqnarray}\label{th10}
R_0&=&|r_0|^2=\dfrac{|r_{01}+r_{12}e^{2i\phi}|^2}{|1+r_{01}r_{12}e^{2i\phi}|^2}\nonumber\\
&=&\dfrac{r_{01}^2+r_{12}^2+2r_{01}r_{12}\cos2\phi}{1+r_{01}^2r_{12}^2+2r_{01}r_{12}\cos{2\phi}}
\end{eqnarray}
The transmittance, or the light intensity transmitted into the Si substrate, can be given by:  
\begin{eqnarray}\label{th11}
T_0&=&1-R_0\nonumber\\
&=&\dfrac{(1-r_{01}^2)(1-r_{12}^2)}{1+r_{01}^2r_{12}^2+2r_{01}r_{12}\cos{2\phi}}.
\end{eqnarray}
Red dashed curve in Fig.~\ref{A_vs_d}b shows the transmittance given by eq.~\ref{th11} as a function of the GaP thickness $d$.  The calculation indicates that the pump intensity transmitted into Si is modified by no more than 15 \% with varying $d$, which is too small to explain the experimental $d$-dependences of $B_\text{Si}$ plotted in the same figure.    

\section{Pump-induced changes in reflectivity from GaP/Si heterointerface}\label{AC}

In this Section we consider the pump-induced change in the probe light wave reflected from the GaP/Si heterointerface. 
We assume that a pump wave induces small disturbances $\delta{n}_1(z,t)$ and $\delta{n}_2(z,t)$ in the refractive indices ${n}_1$ and ${n}_2$ in the GaP layer ($0<z<d$) and in the Si substrate ($z>d$), respectively.  Because the semiconductors have very small or no absorption to the pump light, we can approximate the disturbances to be independent of the depth $z$ in both GaP and Si and to depend only on the time delay $t$ between the pump and probe pulses.  Coherent phonons in the respective semiconductors modulate ${n}_1$ and ${n}_2$ periodically as a function of $t$, whereas photoexcited carriers in Si induces a step-function-like change in ${n}_2$.

To determine the pump-induced change in the reflectance, we can take derivatives of the reflectance with respect to the refractive indices:
\begin{equation}\label{th12}
\dfrac{\Delta R}{R_0}=\dfrac{1}{R_0}\left(\dfrac{\partial R_0}{\partial {n}_1}\delta {n}_1+\dfrac{\partial R_0}{\partial {n}_2}\delta {n}_2\right).
\end{equation}
A change in $n_1$ can modify $r_{01}$, $r_{12}$ and $\phi=n_1k_0d$, whereas a change in $n_2$ can affect only $r_{12}$.  The first and the second terms of eq.~(\ref{th12}) can therefore be expressed by:
\begin{widetext}
\begin{eqnarray}\label{th21}
\dfrac{1}{R_0}\dfrac{\partial R_0}{\partial n_1}\delta n_1&=&\dfrac{1}{R_0}\left(\dfrac{\partial R_0}{\partial r_{01}}\dfrac{\partial r_{01}}{\partial n_1}+\dfrac{\partial R_0}{\partial r_{12}}\dfrac{\partial r_{12}}{\partial n_1}+\dfrac{\partial R_0}{\partial \phi}\dfrac{\partial \phi}{\partial n_1}\right)\delta n_1.\nonumber\\
&=&\dfrac{2(r_{01}^2-1)(1-r_{12}^2)}{(r_{01}^2+r_{12}^2+2r_{01}r_{12}\cos2\phi)(1+r_{01}^2r_{12}^2+2r_{01}r_{12}\cos{2\phi})}\nonumber\\
&\times&\left[\dfrac{r_{01}(1+r_{12}^2)-r_{12}(1+r_{01}^2)}{2n_1}(1-\cos{2\phi})+2k_0dr_{01}r_{12}\sin2\phi\right]\delta n_1\equiv P_1(d)\delta n_1.
\end{eqnarray}
%
\begin{eqnarray}\label{th15}
\dfrac{1}{R_0}\dfrac{\partial R_0}{\partial n_2}\delta n_2&=&\dfrac{1}{R_0}\dfrac{\partial R_0}{\partial r_{12}}\dfrac{\partial r_{12}}{\partial n_2}\delta n_2\nonumber\\
&=&\dfrac{2r_{12}(1-r_{01}^4)+2r_{01}(1+r_{12}^2)(1-r_{01}^2)\cos2\phi}{(r_{01}^2+r_{12}^2+2r_{01}r_{12}\cos2\phi)(1+r_{01}^2r_{12}^2+2r_{01}r_{12}\cos2\phi)}\times\dfrac{-2n_1}{(n_1+n_2)^2}\delta n_2\equiv P_2(d)\delta n_2.
\end{eqnarray}
\end{widetext}
Here we use
\begin{eqnarray}\label{th19}
\dfrac{\partial R_0}{\partial r_{01}}&=&\dfrac{2r_{01}(1-r_{12}^4)+2r_{12}(r_{01}^2+1)(1-r_{12}^2)\cos{2\phi}}{(1+r_{01}^2r_{12}^2+2r_{01}r_{12}\cos{2\phi})^2},\nonumber\\
\end{eqnarray}
\begin{eqnarray}\label{th14}
\dfrac{\partial R_0}{\partial r_{12}}&=&\dfrac{2r_{12}(1-r_{01}^4)+2r_{01}(1+r_{12}^2)(1-r_{01}^2)\cos2\phi}{(1+r_{01}^2r_{12}^2+2r_{01}r_{12}\cos2\phi)^2}.\nonumber\\
\end{eqnarray}
\begin{eqnarray}\label{th20}
\dfrac{\partial R_0}{\partial \phi}&=&\dfrac{-4r_{01}r_{12}(1-r_{01}^2)(1-r_{12}^2)\sin2\phi}{(1+r_{01}^2r_{12}^2+2r_{01}r_{12}\cos{2\phi})^2},
\end{eqnarray}
\begin{eqnarray}\label{th16}
\dfrac{\partial r_{01}}{\partial n_{1}}&=&\dfrac{-2}{(1+n_1)^2}=\dfrac{r_{01}^2-1}{2n_1},
\end{eqnarray}
\begin{eqnarray}\label{th17}
\dfrac{\partial r_{12}}{\partial n_{1}}&=&\dfrac{2n_2}{(n_1+n_2)^2}=\dfrac{1-r_{12}^2}{2n_1},
\end{eqnarray}
\begin{eqnarray}\label{th18}
\dfrac{\partial \phi}{\partial n_{1}}&=&k_0d,
\end{eqnarray}
\begin{eqnarray}\label{th13}
\dfrac{\partial r_{12}}{\partial n_{2}}&=&\dfrac{-2n_1}{(n_1+n_2)^2}=\dfrac{r_{12}^2-1}{2n_1}.
\end{eqnarray}
Equations (\ref{th21}) and (\ref{th15}) reproduce the experimental $d$-dependences of the phonon-induced reflectivity signals from GaP and Si, as shown in Fig.~\ref{A_vs_d}.

\end{document}